\documentclass[aps,pre,reprint,superscriptaddress,floatfix,twocolumn,10pt]{revtex4-2}
\usepackage[english]{babel}
\usepackage{amsmath,bm}
\usepackage{amssymb}
\usepackage{enumitem}
\usepackage{comment}
\usepackage{graphicx}
\usepackage[bottom]{footmisc}
\usepackage{multirow}
\usepackage{array}
\usepackage{ragged2e}
\usepackage{rotating}
\usepackage{keyval}
\usepackage{ifthen}
\usepackage{chemfig}
\usepackage{lmodern} 
\usepackage{mathtools}
\usepackage{tabularx}
\usepackage{braket}
\usepackage{siunitx}
\usepackage{xcolor}
\usepackage{gensymb}

\begin{document}

\title{Reinterpreting ultrafast experiments on supercooled water: Glass transition versus liquid-liquid criticality}

\author{Florian Pabst}
\affiliation{CMSP, The Abdus Salam International Centre for Theoretical Physics (ICTP), 34151 Trieste, Italy}
\author{Ali Hassanali}
\affiliation{CMSP, The Abdus Salam International Centre for Theoretical Physics (ICTP), 34151 Trieste, Italy}

%%%%%% Abstract %%%%%%
\begin{abstract}
Water’s anomalous properties have been hypothesized to originate from a liquid-liquid critical point in the supercooled regime, separating high- and low-density liquid states. Experimental verification remains challenging due to rapid crystallization under these conditions. A recent study reported evidence for such a transition, based primarily on a pronounced increase in the heat capacity of rapidly heated low-density amorphous ice.
Here, we show that this heat capacity increase can be explained without invoking a liquid–liquid transition. By combining simulations using a machine-learning potential trained on the state-of-the-art MB-pol water model, combined with the Tool–Narayanaswamy–Moynihan (TNM) model of the glass transition, we demonstrate that the observed signal can arise instead from a dynamical effect induced by the mobilization of rotational and translational molecular degrees of freedom during ultrafast heating. We further show that our findings are fully consistent with recent electron diffraction measurements showing structural arrest of supercooled water close to our predicted glass-transition temperature. These results provide an alternative interpretation of the experimental observations and highlight the importance of nonequilibrium glassy dynamics in the interpretation of the behavior of supercooled water on ultra-short time scales. 

\end{abstract}

\maketitle

The liquid–liquid transition (LLT) hypothesis was put forward more than 30 years ago based on simulation results \cite{poole1992phase}, following the observation of a first-order-like transition between two glass states, LDA and HDA, observed in experiments \cite{mishima1985apparently}. Although evidence has been collected for the existence of a liquid–liquid critical point (LLCP) in increasingly accurate water models \cite{debenedetti2020second,gartner2022liquid,sciortino2025constraints}, experimental verification remains extremely challenging and disputed \cite{kumar2026rethinking,nilsson2026comment,molinero2026response} due to the proposed location of the LLCP within the crystallization-prone \emph{no man's land}, where experiments on supercooled liquid water are limited to microsecond time scales before crystallization sets in. Nilsson and co-workers have pioneered very challenging experiments making it possible to probe this temperature regime using ultrafast X-Ray experiments \cite{kim2020experimental,you2026experimental}, from which signatures of the presence of an LLT have been inferred. 

Very recently, You et al.~\cite{you2026experimental} reported the observation of a liquid--liquid critical point (LLCP) in supercooled water at 208 $\pm$ 8~K and 1002 $\pm$ 151~bar. Their central evidence is that, upon heating low-density amorphous ice (LDA) with a nanosecond laser pulse of increasing intensity, the temperature rise of the sample saturates beyond a certain energy input. By fitting to a power-law, the authors argue that this saturation reflects a divergence in the near-isochoric heat capacity, from which they infer the location of the proposed LLCP. 

Here, we make the case that the observed increase in heat capacity can also be realized from the glass transition of water on nanosecond time scales. Importantly, our previous simulations~\cite{pabst2026glassy} showed that relaxation times of, for example, collective orientational degrees of freedom, already exceed the nanosecond regime at temperatures where the heat-capacity anomaly was reported in Ref.~\citenum{you2026experimental}. The assumption that water remains equilibrated over the experimental time window therefore requires careful re-examination. To test this possibility, we first establish the connection between the observed heat-capacity increase and the onset of molecular mobility. Subsequently, we examine the concurrent structural evolution of water to probe how these effects are correlated with changes in the heat capacity. Finally, we deploy the Tool–Narayanaswamy–Moynihan (TNM) model to ask whether the heat-capacity response can emerge from purely dynamical effects.

The glass transition is a kinetic phenomenon governed by the ratio of the structural relaxation time $\tau$, to the observation time. Whether a system behaves as an equilibrated liquid or more like a glassy solid therefore depends on its ability to relax within the experimental time window. Conventionally, the glass-transition temperature is defined as the temperature at which $\tau$ reaches approximately 100~s, exceeding the practical time scale of most experiments. In supercooled water within the so-called no-man’s land, however, rapid crystallization restricts accessible observation times to microseconds or less. 

This constraint necessitates extremely rapid temperature ramps, often on nanosecond time scales, which shift the apparent glass transition to temperatures where the temperature dependence of the relaxation time matches the imposed heating rate $q$. This condition can be expressed as~\cite{angell2008insights,kocherbitov2026glass}:
\begin{equation}
    \frac{\text{d}T}{\text{d}t}\frac{\text{d}\tau}{\text{d}T} = q\frac{\text{d}\tau}{\text{d}T}  \approx 1
    \label{eq:deborah}
\end{equation}

From our previous work~\cite{pabst2026glassy}, we have shown that relaxation times exceed the nanosecond regime at temperatures where a divergence in heat capacity was reported in Ref.~\citenum{you2026experimental}. To test whether the experimentally observed increase in heat capacity could instead originate from a glass transition, we performed molecular dynamics simulations using a neural network potential trained on the highly accurate MB-pol water model~\cite{bore2023realistic}. All temperatures from simulations reported here are shifted by the mismatch in melting temperature between simulation and experiment (+10~K) to allow for direct comparison with experimental data.

We closely followed the experimental protocol of You et al. during laser heating: To mimic the annealed LDA used in the experiment, we started from five fully equilibrated independent configurations of 512 water molecules at 220~K (the lowest temperature at which full equilibration is feasible within reasonable time~\cite{pabst2026glassy}). Five additional runs were started from configurations that had been held at 210~K (after equilibration at 220~K) between 7 and 40~ns, where only partial equilibration to an even deeper energy state could be observed. These 10 initial configurations were subsequently quickly (75~K/ns) cooled to 135~K, corresponding to the base temperature of the LDA experiments. Subsequent heating was carried out in the NVT ensemble from these initial configurations to mimic the close to isochoric conditions induced by the laser pulse. 

Because the effective experimental heating rate varies with laser power, we considered two representative heating rates consistent with the reported parameters. As an upper bound, we used a temperature jump of 65~K over a pulse duration of 5~ns, corresponding to 13~K/ns. As a lower bound, we used 4.16~K/ns, based on a somewhat smaller temperature jump of 52.5~K and the delay to the first X-ray probe (12.6~ns). From the internal energy during heating, we computed the heat capacity as $C_V = \text{d}U/\text{d}T$ (see~\cite{pabst2026glassy} for details).

To enable direct comparison with the experimental data of~\cite{you2026experimental}, we converted the reported laser fluence $I_0$ (in J/cm$^{2}$) and temperature increase $\Delta T$ into a heat capacity using the parameters provided by the authors. In particular, we used the distribution of sample thicknesses from their Fig.~S6 to determine the mean thickness $d$, which we corrected for the 20° angle of laser incidence to obtain an effective thickness of $d_{\rm eff} = d/\cos(20\degree) = 35.3~\mu$m$/0.94$. 

The absorbed fraction of the laser energy was estimated using the reported absorption coefficient $\alpha=$~50~cm$^{-1}$ and the Lambert--Beer law, yielding $A = 1-\exp(-\alpha\,d_{\rm eff})=17.1\%$. The density of LDA was taken as $\rho = 0.94$~g/cm$^{3}$~\cite{loerting2011cryoflotation}. Combining these quantities, the experimental heat capacity under isochoric conditions is given by $C_V = A\,I_0/(d_{\rm eff}\,\rho\,\Delta T)$.

In order to correlate the increase of the heat capacity with the onset of a transition between a glass and liquid, i.e., the mobilization of the molecules, we examined the evolution of the translational and rotational dynamics of the system. Specifically, we computed the total electric dipole moment of the box as a proxy for the collective rotational dynamics with a neural network trained to predict the molecular dipoles \cite{malosso2024evidence}, as well as a time-resolved mean square displacement (trMSD),
\begin{equation}
    \text{trMSD}(t) = \frac{1}{N}\sum\limits_{i=1}^{N} |\mathbf{r}_i(t)-\mathbf{r}_i(0)|^2,
\end{equation}
which quantifies the global molecular diffusion of the system as a function of time.

\begin{figure}[h!]
    %\centering
    \includegraphics[width=0.99\linewidth]{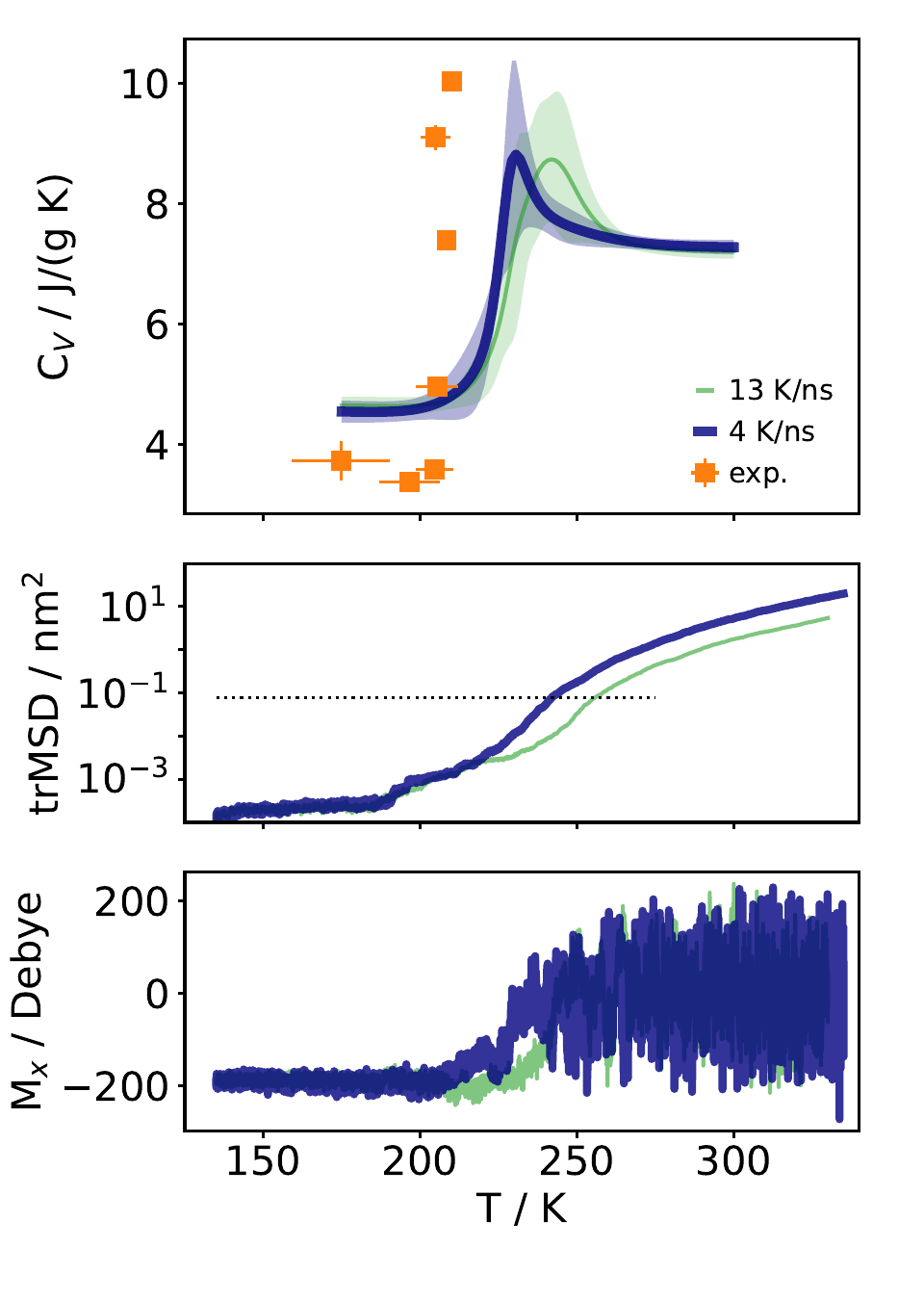}
    \caption{ Comparison between simulated and experimental results. 
\textbf{Upper panel:} Heat capacity traces obtained from molecular dynamics simulations at two different heating rates, together with the experimental data of You et al.~\cite{you2026experimental}. The solid line traces obtained from the simulations indicate an average taken over 10 independent simulations while the shaded colors refer to the standard deviation.  \textbf{Center panel:} Time-resolved mean square displacement (trMSD) from one of the simulation runs. Molecular mobility sets in only above the temperature range corresponding to the heat capacity step, indicating the transition from a glassy to a liquid-like state. The dotted line marks one squared molecular diameter. \textbf{Lower panel:} Total dipole moment in one direction.}
    \label{fig:1}
\end{figure}

Our results are summarized in Fig.~\ref{fig:1}. The simulated heat capacity curves (solid lines are the mean of the 10 runs and shaded areas denote one standard deviation) shown in the topmost panel for both heating rates, exhibit the characteristic features of a glass transition: a step-like increase accompanied by an overshoot. With increasing heating rate, the step shifts to higher temperatures, in agreement with Eq.~\ref{eq:deborah}. Consistently, molecular mobility, quantified via the trMSD (middle panel) and the dipole moment in one direction (bottommost panel) emerges at higher temperatures for the faster heating rate. The dotted line in the middle panel indicates a distance corresponding to one squared molecular diameter, which is reached only after the system has passed through the glass transition. Similarly, in the bottommost panel, the net polarization in the system vanishes at similar temperatures. Below the temperature at which one observes a step-like increase in the the heat capacity there is a net polarization in all three directions which only becomes isotropic after the overshoot. 

Most importantly, the heat capacities obtained from experiment and simulation exhibit very similar features, both in terms of the height and the position of the onset of the growth. Quantitative differences in the exact temperature is rooted either in a too high assumed heating rate in the simulation compared to the experiment or also in addition, missing nuclear quantum effects in the simulation, which are known to shift the glass transition temperature to lower values \cite{eltareb2026nuclear}. Consolidated together, our results point to a different phenomenon underlying the rise in $C_v$: while it was interpreted as a signature of the approach to the LLCP in the experimental work, our simulations suggest that it is intimately connected to a dynamical effect involving the mobilization of the translational and rotational of molecules on ultrashort time scales. 

While the preceding results establish a clear correlation between molecular mobility and the heat-capacity increase, they do not by themselves establish that the observed heat-capacity signal is exclusively dynamical. Unlike simple glass-forming liquids, water undergoes a pronounced structural transformation in the supercooled regime, raising the possibility that the heat-capacity increase may also reflect this structural change. This possibility is particularly relevant here because the structural transformation may occur at the same temperature and over similar time scales as the dynamical glass transition. We therefore examine the structural evolution during ultrafast heating and ask whether it can be distinguished from the dynamical transition identified in Fig.~1.

Recently, isobaric electron-diffraction experiments by Kruger and co-workers have shown that the structure of supercooled water does not evolve further below $200\,\mathrm{K}$~\cite{kruger2023electron}. We therefore extract the position of the first X-ray scattering peak from our simulations during heating and compare it with these experiments, shown in Fig.~2. At low temperatures, the peak position remains constant until it increases to higher values, signaling a densification starting around the same temperature at which the molecules become mobile, as seen in Fig.~1. Interestingly, the structural evolution is largely independent of the heating rate. The overall behavior resembles that observed in the experiment, however shifted to somewhat higher temperatures. This is most likely due to the comparison of isobaric and isochoric data. Indeed, when we examine the evolution of the first diffraction peak for well equilibrated isobaric simulations without any external driving obtained from our earlier study~\cite{pabst2026glassy}, the theoretical predictions are in excellent agreement with the cooling experiments.

The results in Fig.~2 reveal an important complication: in water, the dynamical glass transition is accompanied by a pronounced structural transformation. Because these structural and dynamical changes appear simultaneously, their shared temperature dependence alone cannot tell us directly whether the heat-capacity increase reflects molecular mobility or the structural transformation itself. The fact that the structural evolution is independent from the heating rate, while the heat capacity trace is not, might give a hint that the dynamics play a dominant role. The peak-like overshoot seen upon heating in the heat capacity, by contrast, is a generic feature of glass-forming liquids subjected to an appropriate thermal history \cite{moynihan1974dependence}. The central question is therefore whether the response observed here demands the additional structural physics specific to water, or whether it can emerge from the nonequilibrium dynamics of the glass transition alone.

\begin{figure}[h!]
    %\centering
    \includegraphics[width=0.99\linewidth]{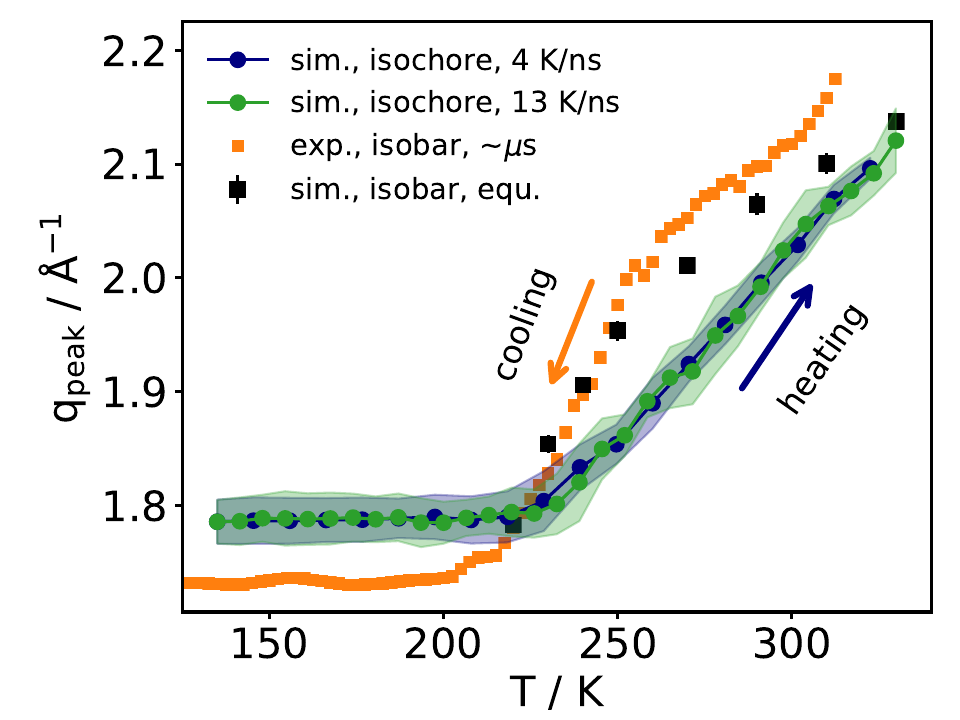}
    \caption{Position of the first diffraction peak ($q_{\rm peak}$) from experiments \cite{kruger2023electron} obtained on cooling on the microseconds time scale and from our simulations upon heating on the nanoseconds time scale (4~K/ns and 13~K/ns), as well as from our previous isobaric equilibrium simulations \cite{pabst2026glassy}.}
    \label{fig:2}
\end{figure}

To test this possibility, we employ the Tool, Narayanaswamy, and Moynihan (TNM) \cite{tool1946relation,narayanaswamy1971model,moynihan1976structural} model, which allows one to express the fictive temperature $T_f$ of a material as a function of the structural relaxation times traversed during the thermal history under consideration. The fictive temperature can be interpreted as the temperature at which a supercooled equilibrium liquid would exhibit the same properties as the out-of-equilibrium glass under consideration. The TNM model can be written as:

\begin{equation}
T_{f,n} = T_0 + \sum_{j=1}^n \Delta T_j \left[ 1- \exp \left[ -\left( \sum_{k=1}^n \frac{\Delta T_k}{q\, \tau_k}\right)^{\beta}\right]\right]
\label{eq:TNM}
\end{equation}

In Eq.~\ref{eq:TNM}, $T_0$ is the starting temperature above the glass transition, from which the temperature is decreased in steps of size $\Delta T$ to below the glass transition and subsequently increased again to $T_0$, with cooling/heating rate $q$. The non-exponential character of the relaxation process is accounted for by the stretching exponent $\beta$, and the relaxation time at step $k$ is given by

\begin{equation}
    \tau_k = \tau_0 \exp\left[ \frac{x\, D}{T_k-T_0} + \frac{(1-x)\,D}{T_{f,k-1}-T_0} \right] \label{eq:tau}
\end{equation}

where a Vogel–Fulcher–Tammann (VFT) temperature dependence is assumed, with parameters $\tau_0$, $D$, and $T_0$. The parameter $x$ is the non-linearity parameter, which determines to what extent $\tau_k$ is governed by the instantaneous temperature $T_k$ or by the fictive temperature $T_{f,k-1}$.

\begin{figure}[h!]
    %\centering
    \includegraphics[width=0.99\linewidth]{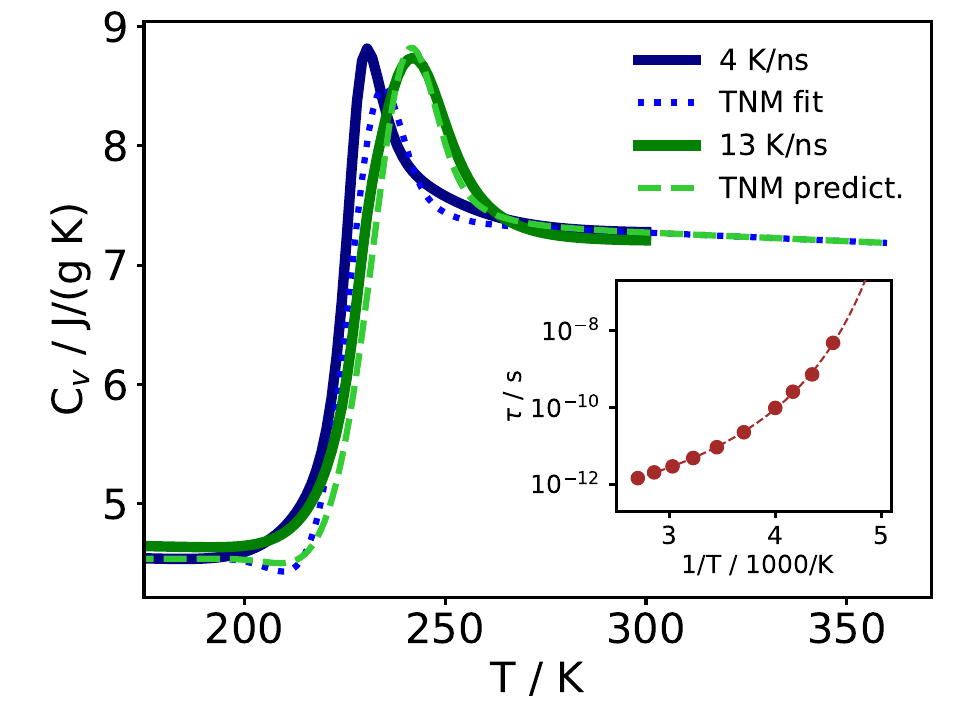}
    \caption{ Comparison of the isochoric heat capacity from our simulation upon heating with 4~K/ns (see Fig.~1) and the fit  with the TNM model, and the heat capacity upon heating with 13~K/ns together with the independent TNM prediction. Inset: Equilibrium molecular dipole relaxation times as a function of inverse temperature with a VFT-fit (dashed line) along the isochore, as used as an input for the TNM model.}
    \label{fig:3}
\end{figure}

In order to use the model to predict the heat capacity of a material subjected to a given temperature history, the temperature dependence of the structural relaxation time must be known, and the values of $x$ and $\beta$ must be determined. It has been shown by comparing dielectric measurements with differential scanning calorimetry experiments that for hydrogen-bonded liquids, the rotational self-correlation time provides a good proxy for the structural relaxation time relevant for calorimetry \cite{wang2008calorimetric,pabst2022understanding}. This quantity is calculated from well equilibrated simulations at different temperatures along the isochore by Fourier transforming the molecular dipole correlation function and using the peak frequency to obtain the relaxation time $\tau = 1/(2\pi f_{\rm max})$, see Ref.~\citenum{pabst2026glassy} for more details. The parameters of Eq.~\ref{eq:tau} are obtained by a VFT-fit to $\tau(T)$ (see inset of Fig.~\ref{fig:3}).

The thermal history employed in the TNM model follows closely the one of the simulation, i.e., full equilibration at 220~K and above, followed by a nanosecond long hold at 210~K, after which the cooling and subsequent heating rate is taken to be identical to that of the simulation.
The heat capacity predicted by the TNM model is given by $C_v = C_v^g + \Delta C_v \frac{\text{d}T_f}{\text{d}T}$, where all nontrivial behavior is contained in $\frac{\text{d}T_f}{\text{d}T}$, i.e., the derivative of eq.~\ref{eq:TNM} with respect to temperature, while $C_v^g$ (the glass heat capacity) and $\Delta C_v$ (the difference between the liquid and glass heat capacities) act merely as scaling parameters and are taken directly from the simulations.
The remaining parameters $x$ and $\beta$ are determined by fitting the TNM model to the heat capacity obtained during the heating simulation with 4~K/ns, which resulted in a non-linearity parameter of $x = 0.85$ and a stretching exponent of $\beta = 0.77$ and a satisfactory agreement between the TNM model and the simulation data as shown in Fig~\ref{fig:3}. 

The structural relaxation---especially of supercooled liquids---commonly exhibits stretched exponential decay, often thought to be influenced by dynamical heterogeneity developing in the supercooled regime. Our value of the stretching parameter $\beta$ is somewhat higher than the value previously found to be a generic value for many supercooled liquids near the glass transition temperature, including salt solutions of water \cite{pabst2021generic,zeissler2025fresh}, but is similar to that of the dipole self-correlation function of our simulations. Our value of the parameter $x$ is higher than the value of 0.635, as reported in a previous simulation study of water \cite{giovambattista2005structural}. Given the large variance in the simulation data underlying the fit, the exact values of $x$ and $\beta$ should therefore not be overinterpreted, but both parameters yield physically reasonable values. Moreover, since all parameters of the TNM model are now determined, the model can be used to predict the heat-capacity trace for the 13~K/ns heating simulation without any adjustable parameters. The excellent agreement between the TNM prediction and the simulation data, shown in Fig.~\ref{fig:3}, demonstrates that the heat-capacity response at a different heating rate originates from a dynamical effect.

In summary, we have provided an alternative interpretation of the experimentally observed increase in heat capacity upon ultrafast heating of low-density amorphous ice. Rather than being a signature of a liquid–liquid transition, our simulations and model calculations show that a heat capacity increase resembling the experimental observation can arise from a glass-transition involving the mobilization of rotational and translational molecular degrees of freedom on the nanosecond timescale. This interpretation also rationalizes earlier experiments from the Nilsson group \cite{you2026experimental} as signatures of a glass transition rather than liquid–liquid coexistence, as we have shown in earlier work \cite{pabst2026glassy}.

Given that the most recent experiments are inherently limited to ultrafast heating to avoid crystallization, it is questionable whether such heating experiments can ever detect equilibrium two-state fluctuations underlying a possible liquid–liquid transition, or whether nonequilibrium dynamics such as those revealed here will inevitably mask them. Experiments involving ultrafast cooling appear more promising, as the anomalous heat capacity of water has, for instance, been detected in this way \cite{kim2020experimental}. Performing such experiments near the purported critical pressure would therefore be interesting to explore, if it ends up being experimentally feasible. 

\vspace{2cm}

\textbf{Acknowledgment}\\
A.H. acknowledges funding from the European Research Council (ERC) under the European Union’s Horizon 2020 research and innovation programme (grant agreement No. 101043272 – HyBOP). The views and opinions expressed are those of the authors only and do not necessarily reflect those of the European Union or the European Research Council Executive Agency. Neither the European Union nor the granting authority can be held responsible for them. \\

%\clearpage

\bibliography{bib.bib}

\end{document}